\documentclass[12pt, article, onecolumn]{IEEEtran}
\usepackage{graphicx}
\usepackage[Algorithme]{algorithm}
\usepackage{times}
\usepackage{algorithmic}
\usepackage{tabularx}
\usepackage{amsmath}
\usepackage{amssymb}
\usepackage{subfigure}

\graphicspath{{figures/}} 

\begin{document}

\title{ Buffer-aware Worst Case Timing Analysis of Wormhole Network On Chip}

\author{\IEEEauthorblockN{Ahlem Mifdaoui and Hamdi Ayed*\footnote{*This work has been done during the master thesis of Hamdi Ayed in 2010 at ISAE} }\\
\IEEEauthorblockA{ University of Toulouse -ISAE \\
ahlem.mifdaoui@isae.fr, hamdi.ayed@isae.fr}
}

\maketitle

\begin{abstract}
A buffer-aware worst-case timing analysis of wormhole NoC is proposed in this paper to integrate the impact of buffer size on the different dependencies relationship between flows, i.e. direct and indirect blocking flows, and consequently the timing performance. First, more accurate definitions of direct and indirect blocking flows sets have been introduced to take into account the buffer size impact. Then, the modeling and worst-case timing analysis of wormhole NoC have been detailed, based on Network Calculus formalism and the newly defined blocking flows sets. This introduced approach has been illustrated in the case of a realistic NoC case study to show the trade off between latency and buffer size. The comparative analysis of our proposed Buffer-aware timing analysis with conventional approaches is conducted and noticeable enhancements in terms of maximum latency have been proved.

\end{abstract}


\section{Introduction}

Among several routing techniques, the wormhole routing  \cite{Dally86}  has become the most implemented routing technique in Networks On Chip (NOC) and multicomputers, and recently in other applications domains like satellites e.g. the Spacewire  \cite{Parkers05} network based on Wormhole routing has been integrated in new generation satellites. This success is incontestably due to its simplicity and its significants interests to make the message latency distance-insensitive in contention-free networks while reducing the storage buffers in intermediate routers.

However, the key argument against using wormhole routing in hard real time context lies in its non deterministic behavior due to possible contentions, which can lead to deadlock situation e.g. no message can be transmitted because all the system queues are busy. Hence, achieving a real time behavior with low latency over wormhole based networks still needs the use of additional mechanisms. Various solutions are recently offered to overcome these limitations as the integration of virtual channels allowing the messages to bypass each other in case of conflict, and the implementation of deadlock-free routing algorithms based generally on deterministic approaches, such as Round Robin.

Even-though these solutions eliminate the deadlock when using the wormhole routing, there are not sufficient to prove that traffic deadline constraints are effectively respected to fulfill the real time applications requirements. This may occur when simultaneous traffic flows attempt to share the same resource which can lead to a chain blocking situation e.g. a blocked packet can block another packet which in turn blocks other packets, and so on. Hence, the real-time packet schedulablity has to be proved and the response-time based schedulablity tests are usually used in such a distributed system, where the critical resource is not the computational power but the transmission medium bandwidth utilization. In Wormhole routing networks, this latter depends mainly on flows contentions scenarios, which make the exact Worst Case Response Time (WCRT) calculus very complex due to the huge possibilities of flows arrivals and blocking. In order to handle this problem, many approaches have been proposed in the literature to compute the maximum WCRT for wormhole NoC using scheduling theory  \cite{Balakrishnan98, Lu05, Burns08} and Network Calculus  \cite{Qian10, Thomas11}. However, none of these methods integrate the buffer's size impact on WCRT bounds.


Hence, our main contributions in this paper are three hold. \textbf{First}, the analysis of the router buffer's size and the packets lengths impact on the different dependencies relationship between flows is conducted, and more accurate definitions of direct and indirect blocking flows sets \cite{Kim98} are introduced. \textbf{Second}, a buffer-aware worst-case timing analysis of wormhole networks on chip is defined by taking into account the newly defined direct and indirect blocking flows sets, based on Network Calculus formalism \cite{Leboudec}. \textbf{Third}, this introduced approach is illustrated in the case of a realistic Network On Chip application and the obtained results are compared to those obtained with conventional methods. A noticeable enhancement of WCRT bounds has been proved.  

In the next section, we review the most relevant works to provide worst case performance analysis of Wormhole routing networks and relate them to our work. Afterwards, the worst-case buffer-aware timing analysis of the Wormhole NoC is tackled as follows. First, the identified limitations of the current definition of interrelationships between traffic flows, and the proposed enhancements to integrate the buffer's size and packets lengths impact are detailed in section 3. Then, Section 4 and Section 5 show the Wormhole NoC modeling and the buffer-aware worst-case timing analysis based on Network Calculus formalism, respectively. Finally, its practical feasibility and comparative analysis with conventional approaches are illustrated in section 6. Section 7 concludes the paper.


\section{Wormhole NoC and Real Time}


The Wormhole routing has been proposed for NoC \cite{Dally86} to reduce the implemented memory in intermediate routers and bridge the gap between Virtual Cut Through and Circuit routering techniques \cite{Lionel93}. The Wormhole routing is a kind of Cut Through where the packet is divided into a fixed size flits (commonlyt one Byte). The header flit contains the route information and it is transmitted along the identified route. As long as the header advances along the network, the rest of the packet flits follow in a pipeline way. If the header flit is blocked because of a network contention, rather than buffering the entire packet in one intermediate node as Virtual Cut Through, the trailing flits remain in the routers along the established path which is less memory consuming; and unlike the Circuit switching, the physical circuit is not turned off thanks to blocking the packet header in the current crossed node which allows the physical channel sharing and enhances throughput.

However, with the Wormhole routing, as several packets try to access the same physical channel simultaneously, contentions may occur and can lead to deadlock situation. One way to address this problem is to split each physical communication channel into several virtual channels where each one has its own buffer and flow control allowing the messages to bypass each other in case of conflict. However, the main issue concerning this solution is the accurate design of the virtual channels multiplexing technique to maximize the physical channel utilization. A more common solution lies in implementing a deadlock-free routing algorithm. There are two kinds of routing algorithms: (i) the first ones are simple to implement and are based on a deterministic approach to avoid cycles in the channel dependence graph, such as Round Robin Algorithm; (ii) the second can react dynamically to network conditions and use an adaptive approach to enhance throughput but require sophisticated hardware and increase the system's complexity. In this paper, only the deterministic wormhole routing are considered and the impact of buffer size within routers on the timing performance is detailed. 

Several timing analysis studies have been proposed in the literature for wormhole NoC. These approaches range from predicting average flows latencies using  simulation like  \cite{Ramany94, Lionel97} and concepts from queuing theory like \cite{Guz06, Hu06, Ogras07, Arjomand10}, to defining worst case latencies using scheduling theory like  \cite{Balakrishnan98, Lu05, Burns08}  and Network Calculus like  \cite{Qian10, Thomas11} . Since the idea in this paper consists in studying the worst case timing analysis of Wormhole NoC to fulfill the real time applications requirements, we review in this section the most relevant works based on deterministic approaches, e.g. scheduling theory and Network Calculus.

In order to verify the packet delivery feasibility of Wormhole routing NoC, some researchers used the scheduling theory to find the worst case delay bounds for traffic flows. The idea of \cite{Balakrishnan98} was based on \textit{lumping} all the crossed links for each flow to be considered as one shared resource to calculate the worst case latency. This approach is quite simple but pessimistic because it did not take into account the different kinds of flows inter-relationships, i.e., direct and indirect blocking. Unlike this approach, \cite{Lu05} considered  a more accurate model with the \textit {contention tree}, which reflects the different network contentions, and consequently the obtained worst-case delays are less pessimistic than in \cite{Balakrishnan98}. However, this approach needs a static bandwidth partitioning method for real time and non real time messages and a global ordering messages, which leads to an increasing resolution's complexity for large-scale networks. In \cite{Lu05} and \cite{Burns08}, authors  distinguished direct and indirect contentions for each traffic flow. Then, the existent worst case response time calculus in scheduling theory has been extended to wormhole routing NoC, by integrating the indirect interferences impact as a jitter to obtain a recursive formulae to calculate the total worst case latencies. The benefits of this approach are interesting for small packets compared to buffer size within crossed routers. However, if it is not the case, the obtained delays could be unacceptable. In addition, the example considered in \cite{Burns08} did not reveal the recursive calculus complexity to handle a large-scale network with a high amount of traffic. 

Recently, there are some interesting works to evaluate the worst case timing performance of Wormhole routing networks using the Network calculus formalism. In \cite{Qian10}, the authors modeled the different characteristics of wormhole NoC using end to end service curves, where the feedback flow control mechanism was integrated as a virtual controller of buffer's limitation at each input port. Then, after analyzing the different flow interferences patterns, maximal delay bounds were recursively deduced using network calculus results. This method leads to a complex system model due to a lot of dependencies generated by the flow control model, which makes its use limited to small-scale networks. Another paper \cite{Thomas11} deals with the same problem in the specific case of the Spacewire technology \cite{Parkers05}, which did not integrate the buffer size impact on WCRT bounds. 

All these approaches did not take into account the router buffer's size and the packet length impacts on the interferences between flows, and consequently timing performance. Hence, our main idea in this paper is to introduce a buffer-aware timing analysis of wormhole NoC. First, like \cite{Kim98} and \cite{Burns08} approaches, the direct and indirect interferences are considered for each flow, but in a more accurate way when integrating the buffer's size and the packets lengths impacts. Then, based on Network Calculus formalism, Wormhole NoC is modeled and timing analysis is conducted. Finally, a realistic NoC case study is considered to show the trade off between latency and buffer size, and comparative analysis of our proposed Buffer-aware timing analysis with conventional approaches is illustrated and noticeable enhancements in terms of maximum latency have been proved.


\section{Direct and Indirect Interferences under Wormhole Routing}

\subsection{ Definitions and Assumptions}
Notations described in \textbf{table \ref{tab1}} are used in the rest of the paper. 
\begin{table}[h!]
\caption{Notations}
\label{tab1}
\begin{center}
\begin{tabularx}{\linewidth} {l X}
\hline
$C$ & links transmission capacity \\
$\epsilon$ & Technological router relaying latency \\ 
$Buff$ & the router input port buffer size \\
$F$ & the traffic flows set sent by all the nodes on the networks \\
$n_{k}$ & routers number in $path_k$ \\
$path_k$ & the path of the flow $k$ consisting of the set of crossed network components \\
$path_k(n)$ & the nth crossed node on the $path_k$ \\
$path_k(i \to  n)$ & the subpath of flow $k$ between the ith and the nth crossed node on the $path_k$ \\
$subpath_i^k$ & the subpath of the flow $i$ from the last physical intersection point between the two flows $i$ and $k$ until their real divergence point due to flow $i$ packet length and buffer size\\
$F_{DB}^{k}$ & the traffic flows set imposing a direct blocking delay to traffic flow $k$ \\
$F_{IB}^{k}$ & the traffic flows set imposing an indirect blocking delay to traffic flow $k$ \\
$O_R(k)$ & the associated output port for flow $k$ in router $R$ \\
$I_R(k)$ & the associated input port for flow $k$ in router $R$ \\
$F_l^p$ & the traffic flows set sent from the input port $p$ to the output port $l$ \\
$L_{max}(F)$ & the maximum packet length belonging to traffic flows set $F$ \\
\hline
\end{tabularx}
\end{center}
\end{table}

The traffic schedulablity is analyzed in this paper using response time based schedulablity test and Network Calculus formalism. To handle the complexity of the exact Worst Case Response Time (WCRT) Calculus, an upper bound of this latter is considered herein and compared to the respective deadline. However, this schedulablity test results in a sufficient but not necessary condition due to the pessimism introduced by the upper bounds. Nevertheless, we can still infer the traffic schedulablity by comparing the computed WCRTs with the respective deadlines, i.e., \\
\begin{center}
$\forall k \in messages$ , WCRT$_k \leq Deadline_k  \Longrightarrow $ The messages set $messages$ is schedulable 
\end{center}

The upper bound of $WCRT_k$ associated to traffic flow $k$ crossing the network corresponds to its maximal end to end delay bound from its source to its destination and it consists of two parts:
\begin{itemize}

\item in case of contention-free network, the traffic flow $k$ is sent alone on the network and its end to end communication latency depends on its  $path_k$ and its packet size $L_{max}^k$. This part is the minimum end-to-end transmission delay $D_{TR}^{k}$ and it is as follows:
\begin{equation}
	\label{DTR}
	 D_{TR}^{k} = \frac{L_{max}^k}{C} + n_k . \epsilon
\end{equation} 

\item In case of conflicts, the traffic flow $k$ can be disturbed and its maximum latency is increasing due to the different types of interferences, which will be detailed in the next section. This blocking delay is called $D_{B}^k$.
\end{itemize}

Hence, The maximal end to end delay communication bound of a given flow $k$ is as follows:
\begin{equation}
	\label{Deed}
	 D_{eed}^{k} = D_{TR}^{k} + D_{B}^k
\end{equation} 

Moreover, the schedulabilty test becomes as follows:
\begin{center}
$\forall k \in messages$, $D_{eed}^{k} \leq Deadline_k  \Longrightarrow $ The messages set $messages$ is schedulable 
\end{center}

\subsection{Conventional approaches and identified limitations}
In order to evaluate the maximal network latency bounds,  the authors in \cite{Kim98} and \cite{Burns08} consider the different dependencies between flows with a blocking dependency graph for each message stream. This graph shows two kinds of interferences between traffic flows:
\begin{itemize}
\item \textit{Direct Blocking}: for a given flow, this interference is due to all the flows that have at least one physical link in common with this latter and respect the priority based mechanism if any. Hence, with an arbitration priority policy within routers, all traffic flows with higher priority than the considered one and at maximum one maximum packet length of lower priority can cause this kind of interference.
\item \textit{Indirect Blocking}:  this interference is caused by all the traffic flows that do not share any physical link with the considered flow but have at least one physical link with at least one traffic flow leading to a direct blocking.   
\end{itemize}
\begin{figure}[h]
\centering
\includegraphics[scale=0.9]{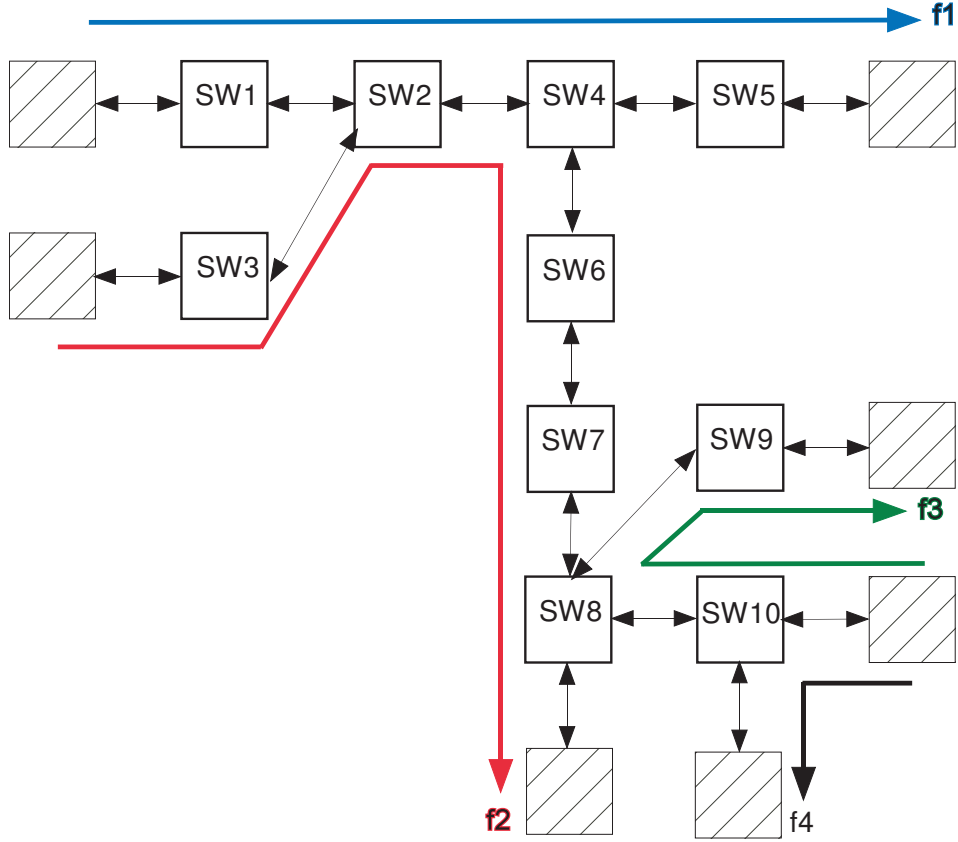}
\caption{\it An example of Wormhole NoC}
\label{example}
\end{figure} 
Based on the example illustrated in Fig. \ref{example}, we identified some limitations of these conventional definitions, which do not integrate the impact of input port buffer size in routers. The determination of indirect interferences is a major step to evaluate the worst case latency for a considered flow and these interferences depend on the buffer size of each crossed input port. In fact, in the illustrated example, if we consider an input buffer size about $56$ Bytes within each router, one can notice that the flow $f2$ with a packet length $100$ Bytes will need just $2$ hops to completely liberate the output port of router $4$. Hence, when $f2$ reaches router $8$ the considered flow $f1$ will not be blocked any more by $f2$. Hence, the flows that interfere with the flow $f2$ after this divergence point, like the flow $f3$, will not induce an indirect blocking on $f1$ any more. As one can notice, the size of the buffer within routers is extremely important to optimize the flows set leading to indirect blocking. Hence, the impact of the buffer size and the packet lengths have to be integrated to determine the different flows sets leading to direct or indirect blocking. 

%
%

\subsection{Proposed enhancements}
In order to handle the different identified limitations, we introduce a more accurate inter-relationships between traffic flows to integrate the impact of the input port buffer size and the packets lengths.

\subsubsection{Direct Blocking flows Map}
The traffic flows set imposing a direct blocking delay to traffic flow $k$ is currently defined as $F_{DB}^{k} = \{i \in F ,\ path_i \bigcap path _k \neq \o\}$. To integrate the identified impact parameters like buffer size and packet length, we define the following  sub path  between the considered flow $k$ and each flow in the direct blocking flows set:
\begin{equation}
\label{subpath}
subpath_i^k = path_i(Last_i^k \to Divergence_i^k)
\end{equation} 
Where
\begin{displaymath}
\left\{ \begin{array}{l}
Last_i^k = \max\{n \in N^*/ \ path_i(n) \in path_k\} \\
Divergence_i^k = Last_i^k + hops_i\\
hops_i = \lceil \frac{L_i}{Buff} \rceil \\
\end{array} \right . 
\end{displaymath}

This new parameter will be integrated to define the direct blocking flows map for a considered flow, which defines the associated subpath for each flow that induces a direct blocking to traffic flow $k$. The obtained relationship is as follows:
\begin{eqnarray}
Map_{DB}^{k} &:& F_{DB}^{k} \to E
\nonumber\\
& & f_l \to  subpath_l^k 
\end{eqnarray}

\subsubsection{Indirect Blocking flows Map}
The traffic flows set imposing an indirect blocking delay to traffic flow $k$ is currently defined as $F_{IB}^{k} =\{i \in F \setminus F_{DB}^{k}, \exists l \in F_{DB}^{k},\  path_i \cap path _l \neq \o \}$. This definition is enhanced by introducing the subpath notion defined bellow. In order to optimize this flows set, we consider only the traffic flows that do not share any physical link with the considered flow $k$ but have at least one physical link with one traffic flow $l$ leading to a direct blocking on the $subpath_l^k$. Hence, the accurate definition of indirect blocking flows is:
\begin{eqnarray}
	\label{FIB}
	F_{IB}^{k} &= & \{i \in F \setminus F_{DB}^{k},\ \exists l \in F_{DB}^{k},\  	\nonumber\\
				  & & path_i \cap subpath_l^k \neq \o \}
\end{eqnarray}  

The indirect blocking flows map for a considered flow $k$ defines for each flow that induces an indirect blocking delay, its associated subpath with one of the flows in $F_{DB}^k$:
\begin{eqnarray}
Map_{IB}^{k} &:& F_{IB}^{k} \to E
\nonumber\\
& & f_i \to  Map_{DB}^{l} (f_i) / \ l \in F_{DB}^{k}
\end{eqnarray}

Hence, one can see that when buffer size increases, 
the $F_{IB}^{k} \to \o$.



\section{Wormhole Routing Modeling}

\subsection{Network Calculus Concepts}
To evaluate the QoS level offered by the network, the maximal end to end delay bounds will be compared to the temporal deadlines. To achieve this aim, we have chosen to conduct analytic studies instead of simulations, which are commonly used to validate models. In fact, simulations cannot cover the entire domain of the model applicability and specially rare events that represents worst-case functioning. Moreover, these latter are always conducted with a given confidence level always less than 100 percent. So, clearly, simulations cannot provide the deterministic guarantees required by our critical application, where a failure might have a disastrous consequence on our system.

Our analytic study is based on the use of \emph{Network Calculus} theory, introduced by Cruz \cite{Cruz1} and developed in a neater way by Leboudec \cite{Leboudec}, because it is well adapted to controlled traffic sources and provides deterministic end-to-end delay bounds. This formalism \cite{Leboudec} is based on \textit{min-plus} algebra for designing and analyzing deterministic queuing systems where the compliance to some \emph{regularity constraints} is enough to model the traffic. These constraints limit traffic burstiness in the network and are described  by the so called \emph{arrival curve} $\alpha (t)$, while the availability of the crossed node is described by a \emph{service curve} $\beta (t)$. The knowledge of the arrival and service curves enables the computation of the delay bound that represents the worst case response time of a message, and the backlog bound that is the maximum queue length in the node. The delay bound $D$ is the maximal horizontal distance between $\alpha (t)$ and $\beta (t)$ whereas the backlog bound $B$ is the maximal vertical distance between them. 

This formalism gives an upper bound for the output flow $\alpha^*(t)$, initially constrained by $\alpha(t)$ and crossing a system with a service curve $\beta(t)$, using min plus deconvolution $\oslash$ where:
\begin{equation}
	\label{Deconvolution}
	 \alpha^*(t) = sup_{s \geq 0}(\alpha(t+s) - \beta(s)) = (\alpha \oslash \beta)(t)
\end{equation} 
	

Another important result given in the Network Calculus formalism is the concatenation theorem that is as follow: 

\textit{Assume a flow with arrival curve $\alpha(t)$ traverses systems S1 and S2 in sequence where S1 offers service curve $\beta1(t)$ and S2 offers $\beta2(t)$. Then, the concatenation of these two systems offers the following single service curve $\beta(t)$ to the traversing flow:}
\begin{equation}
	\label{PBOO}
	\beta (t) = (\beta1 \otimes \beta2) (t) = inf_{0 \leq s \leq t} (\beta1(t-s) + \beta2(s)) 
\end{equation}

There is also another known result concerning the blind multiplexing:

\textit{Assume flows 1 and 2 with arrival curves $\alpha1(t)$ and  $\alpha2(t)$ traverse system S which offers a strict service curve $\beta(t)$. Then, the minimal service curve offered to flow 1 is: }
\begin{equation}
	\label{BM}
	\beta1 (t) = \left( \beta(t) - \alpha2(t) \right)^{+}
\end{equation} 

where the notation $ x^{+} = \max(0,x)$

\subsection{Traffic Model}
\label{traffic}
Four parameters $(T_i, D_i, L_i, J_i)$ are defined for each traffic flow $i$:
\begin{itemize}
	\item The periodicity $T_i$: for a periodic message, it is the period and for a sporadic message, it is low bounded as its minimal inter-arrival time.
	\item The temporal local deadline $D_i$: (the message life duration) it is the period for a periodic message and the maximal response time for a sporadic message.
	\item The length $L_i$: the maximum length of a message 
	\item The jitter $J_i$: the maximum deviation of successive packets arrivals.
\end{itemize}

Hence, each traffic flow $i$ has an affine arrival curve $\alpha_i$:
\begin{equation}
	\label{arrival}
	\alpha_i (t) = L_i + \frac{L_i}{T_i}(t + J_i)
\end{equation} 

\subsection{Wormhole Router Model}
\label{router}
The most important characteristics of wormhole routers are as follows:

\begin{itemize}
	\item the Weighed Round Robin scheduling is used to share the link bandwidth between packets;
	\item the path of each flow is statically determined thanks to a static routing table in each router;
	\item routers contain one finite size buffer per input port and no buffer in the output port;
	\item each router sends back a control information to the upstream router to indicate the state of the input buffer (free or not).
\end{itemize}

\begin{figure}[h]
\centering
\includegraphics[scale=0.9]{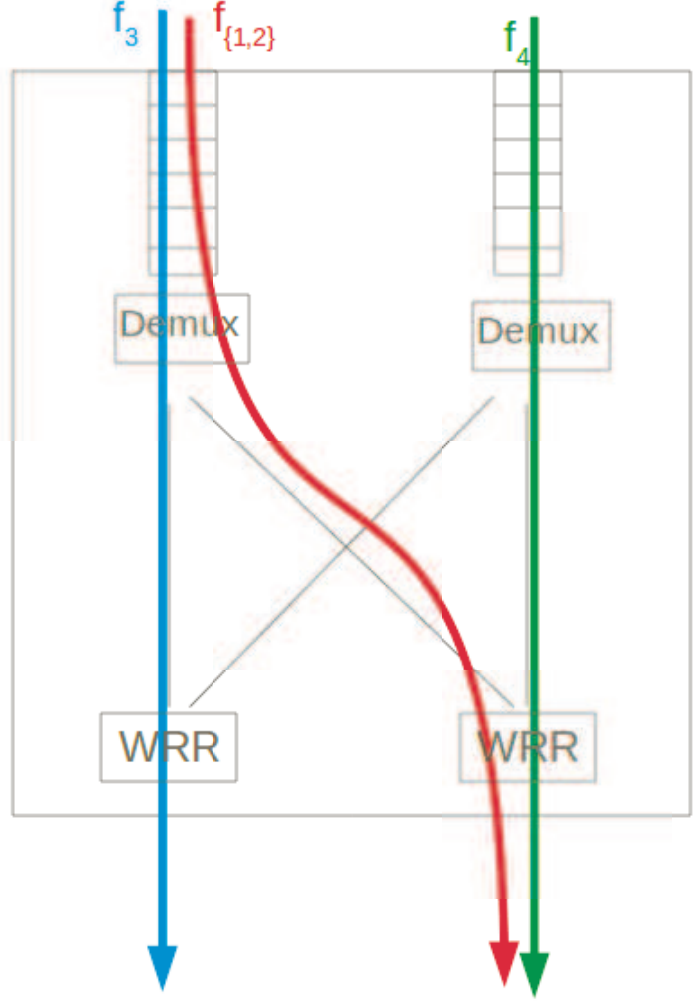}
\caption{\it Wormhole router model}
\label{ModelSPW}
\end{figure}

These characteristics are taken into account when modeling the wormhole router, as shown in the figure \ref{ModelSPW}.

The finite buffer in input ports and the absence of buffers in the output complicates the analytical model. In fact, without control feedback and finite size buffers, buffer overflow is unavoidable. In order to integrate theses constraints, we consider an iterative method to compute the indirect blocking delay experienced by each flow and its impact on the end to end delay.  


On the other hand, routers perform Weighted Round Robin (WRR) scheduling to serve traffic flows trying to access the same output port. With the Weighed Round Robin scheduling, the flows are served according to their weights. Hence, for each router output port $l$, an appropriate weight $\phi_{l,F_l^p}$  is considered for each traffic flows set $F_l^p$, received by the output port $l$ from input port $p$ in the router $R$. This weight respects the stability conditions $\sum_p \phi_{l,F_l^p} = 1$ and $\forall i \in F_l^p, \ \frac{L_i}{T_i} \leq \phi_{l,F_l^p}. C$ in each router. To determine the service curve offered by the WRR node, the Leboudec's result \cite{Leboudec} concerning the service curve offered by a Generalized Processor Sharing (GPS) node is used. The researched service curve $\beta_{l,F_l^p}$ offered by the router output port $l$ to the traffic flows set $F_l^p$ sharing the same input buffer $p$ is then:

	\begin{equation}
	\label{WRR}
	\beta_{l,F_l^p}(t) = c_{l,F_l^p}(t - \epsilon) 
	\end{equation}

Where $c_{l,F_l^p} = \phi_{l,F_l^p}.C$ and $\phi_{l,F_l^p}=\frac{r_{F_l^p}}{\sum_{k} r_{F_l^k}}$.\\

Another effect that has to be taken into account, is the blocking of a flow in the input buffer by other flows that don't share the same output port. As we can see in the example in the figure \ref{ModelSPW}, the flow $f_{1,2}$ has to wait the end of transmission of $f_3$ by its associated output port. We will refer to this effect the demultiplexing effect.

\begin{figure}[h] 
      \centering \includegraphics[scale=0.9]{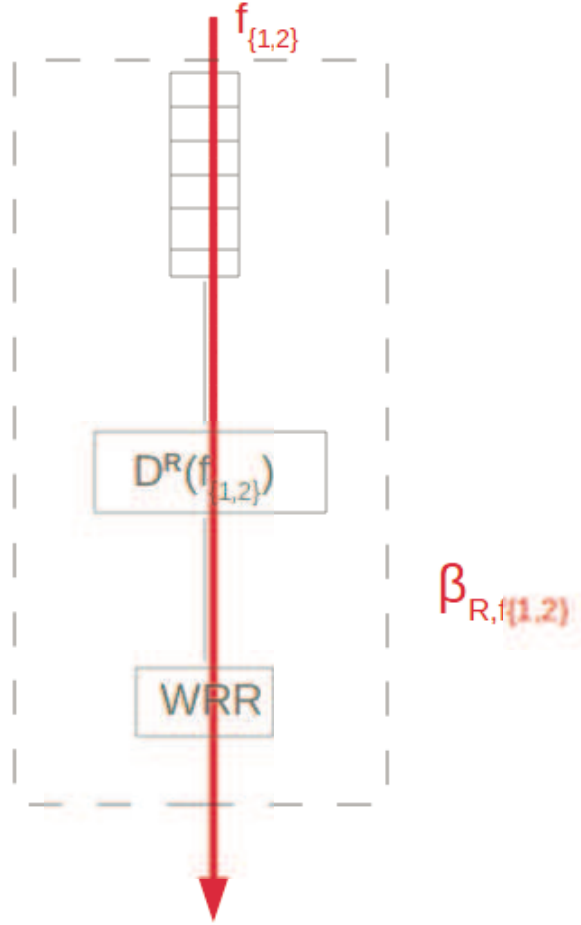}
      \caption{\it router service for $f_{1,2}$}
       \label{DemuxEffect}
\end{figure}

As shown in the figure \ref{DemuxEffect}, we take into account the demultiplexing effect on the aggregated flow $f_{1,2}$ by integrating a Dirac service curve $\delta_{D^R(f_{1,2})}$ that represents the worst case delay that $f_{1,2}$ undergo due to transmission of flow $f_3$ from its associated output port. 

Due to the demultiplexing effect a flow $k$ is delayed in the input buffer by the flows sharing the same buffer and going for a different output ports in the router $R$, this set is $\cup_{l\neq O_R(k)}{F_{l}^{I_R(k)}}$ (we consider only the flows in direct blocking flows set $F_{DB}^k$). The flow of interest have to wait until these flows are transmitted. Hence, the aggregated flow $F_{O_R(k)}^{I_R(k)}$ is delayed, in the worst case, in the router $R$ by:

\begin{equation}
D^R(k)=\sum_{l \neq O_R(k)}\sum_{j \in F_l^{I_R(k)}}{\frac{b_j^R}{C}}
\end{equation}

where $b_j^R$ is the arrival burst of flow $j$ in the router $R$ and $\beta_{l,F_l^p}$ the service for the traffic flows set $F_l^p$ coming from port $p$ to the output port $l$. Hence, the service guaranteed by the router to the aggregated flow going out from a giving output port and coming from a giving input buffer is:

\begin{equation}
\beta_{l,F_{l}^{I_R(k)}}(t)=c_{l,F_{l}^{I_R(k)}}(t - \epsilon) \otimes \delta_{D^R(k)}
\end{equation}

The individual router service curve offered by router $R$ to one traffic flow $k$ depends on the arrival curves of the rest of the traffic $F_{O_R(k)}^{I_R(k)}$ sharing the same input buffer in router $R$ and going to the same output port. It's is obtained using the blind multiplexing property (\ref{BM}): 

\begin{equation}
\label{IRR}
	\beta_{R,{k}}(t) = \left( \beta_{O_R(k),F_{O_R(k)}^{I_R(k)}}(t) \otimes \delta_{D^R(k)} - \sum_{j \in F_{O_R(k)}^{I_R(k)} \setminus \{k\}} {\alpha_j^R} \right)^{+}
\end{equation}

where $\alpha^R_j$ is the arrival curve of the individual flow $j$ to the router $R$ described in section \ref{traffic} by resolving the burstiness constraint evolution from one crossed router to another using equation (\ref{Deconvolution}).


\section{Buffer-Aware Worst Case Timing Analysis}
In this part, we explain the calculation of the maximal end to end delay bound for a traffic flow in a wormhole NoC, using the Network Calculus formalism \cite{Leboudec}, by integrating the impact of the buffer size within the routers.

\subsection{Direct Blocking Delay Bounds}
In order to calculate the direct blocking delay, we derive service curves for individual flows in each crossed router, then we get the end to end service curve using the concatenation.

Thanks to the concatenation theorem (\ref{PBOO}) and the integration of the individual service curve in each crossed router given by (\ref{IRR}), the end to end service curve offered to an individual traffic flow $i$ is:
\begin{equation}
\label{EEDS}
	\beta_{i}(t) = \otimes_{R \in path_i} \beta_{R,{i}}(t)
\end{equation}

The sum of the transmission and direct blocking delay bound for each traffic flow $i$ is the maximal horizontal deviation between  the arrival curve of the packet $\alpha_{i}$ and $\beta_i$, where:

\begin{equation}
\label{alpha}
	 D_{DB}^{i} = h(\alpha_{i}, \beta_{i}) 
\end{equation}

\subsection{Indirect Blocking Delay Bounds}
The indirect blocking delay of a traffic flow $k \in F$ is as follows:
\begin{equation}
	\label{DIB}
	 D_{IB}^{k} = \sum_{i \in F_{IB}^k } D_{DB}^i (Map_{IB}^k(i)) + D_{IB}^i
\end{equation} 

where:

\begin{eqnarray}
\label{DDB}
D_{DB}^{i} (Map_{IB}^k(i)) & = & h(\alpha_{i}^{(Map_{IB}^k(i))(0)}, \beta_{i}(Map_{IB}^k(i)))  \nonumber \\
\beta_{i}(Map_{IB}^k(i)) & = & \otimes_{R \in Map_{IB}^k(i)} \beta_{R,{i}}(t) \nonumber \\
\end{eqnarray}


With $\alpha_{i}^{(Map_{IB}^k(i))(0)}$ is the arrival curve of the flow $i$ at the input of the first router on subpath $Map_{IB}^k(i)$. 

The traffic flow $k$ has to wait in the worst case the transmission of the indirect blocking traffic flows set which consists of direct and indirect blocking delays. The different direct blocking delays are calculated thanks to (\ref{DDB}). Then, in order to calculate the indirect blocking delay, we proceed with a recursive calculus that requires the execution of the following algorithm \ref{alg1}. This algorithm initially identifies the flows sets imposing the direct blocking and indirect blocking delays to the flow of interest $f_k$ (lines $2-3$). The new flow set $F^*$ is considered to ignore the contention interference due to $F_{DB}$ when calculating the indirect blocking delay (line $5$). We have to recursively compute the indirect blocking delay of each flow in $F_{IB}$ by reducing in each loop the flow set $F^*$ until it is equal to null. This means that the considered flow does not suffer from indirect blocking and its direct blocking delay is equal to zero (lines $6-8$). This algorithm has a complexity about $\mathcal{O}(card(F)^2)$.

\begin{algorithm}
\caption{Indirect blocking delay bounds calculus }
\label{alg1}
\begin{algorithmic}[1]
\STATE ComputeIndirectBlockingDelay ($f_k$, $path$, $F$)
\STATE $F_{DB} \leftarrow F_{DB}^k$
\STATE $F_{IB} \leftarrow F_{IB}^k$
\STATE $Map_{IB} \leftarrow Map_{IB}^k$
\STATE $F^* \leftarrow F \setminus F_{DB} $
\STATE IF ($F^* == \o ~ or ~ F_{IB} == \o $) $Delay \leftarrow 0 $
\STATE ELSE $Delay \leftarrow \sum_{f_l \in F_{IB}} D_{DB}^l(Map_{IB}(f_l) +$ ComputeIndirectBlockingDelay($f_l$, $Map_{IB}^k(f_l)$, $F^*$)
\STATE RETURN $Delay$
\end{algorithmic}
\end{algorithm}


\section{Performance Evaluation}

\subsection{Case study}
The considered case study is the same than the one considered in \cite{Thomas11} in terms of traffic and architecture. However, the buffer size within the routers is varied to analyze its impact on the maximum latency, and it is ranging between $1$Bytes and $1000$Bytes. 

\subsection{Analytical delay bounds vs buffer size}

The maximum latency for each buffer size value has been computed based on our detailed model and the proposed buffer-aware timing analysis. This computation has been conducted based on WoPANets tool \cite{Wopanets}.  Enhanced bounds have been obtained, compared to the results in \cite{Thomas11}, when the buffer size is at least equal to 20 bytes, and an amelioration of at least 30\% has been shown when the buffer size increases. 


\section{Conclusion}
A buffer-aware worst-case timing analysis of wormhole NoC has been proposed in this paper to integrate the impact of buffer size on the different dependencies relationship between flows, and consequently the timing performance. Hence, more accurate definitions of direct and indirect blocking flows sets have been introduced. Afterwards, the modeling and worst-case timing analysis of wormhole networks on chip have been detailed, based on Network Calculus formalism. This introduced approach has been illustrated in the case of a realistic Network On Chip application and the obtained results are compared to those obtained with conventional methods. A noticeable enhancement of WCRT bounds has been proved when the buffer size increases.



\bibliographystyle{IEEEtran}
\bibliography{mifdaoui-rts-bib}

\end{document}